# Ethical Aspects of Internet of Things from Islamic Perspective

Wazir Zada Khan, Mohammed Zahid, Mohammed Y Aalsalem, Hussein Mohammed Zangoti and Quratulain Arshad

Farasan Networking Research Laboratory, Faculty of Computer Science and Information System, Jazan University, Kingdom of Saudi Arabia

{wazirzadakhan, mzahid, m.aalsalem, hmzangoti}@jazanu.edu.sa, brightsuccess_12@yahoo.com

*Abstract*— The Internet of Things (IoTs) is an evolving new face of technology that provides state of the art services using ubiquitously connected smart objects. These smart objects are capable of sensing, processing, collaborating, communicating the events and provide services. The IoT is a collection of heterogeneous technologies like Sensor, RFID, Communication and nanotechnology. These technologies enable smart objects to identify objects, collect information about their status, communicating the collected information for taking some desired actions. Widespread adaptations of IoT based devices and services raised the ethical challenges for their users. In this paper we highlight ethical challenges raised by IoT and discuss the solutions and methods for encouraging people to properly use these technologies according to Islamic teachings.

## I. INTRODUCTION

The Internet of Thing (IoT) connects multiple smart objects to the Internet with the purpose of capturing, and communicating data and providing outstanding services, ultimately improving the quality of human life. Some of the compelling applications of IoT can be seen in healthcare, industrial automation, transportation etc. A large number of SMART IoT applications can be categorized into three main classes; Individual based, Community/Environmental based and Enterprises/Industrial businesses based (also shown in Fig. 1) [1-5].

The number of physical smart objects in IoT is increasing at a high rate and according to Cisco report by 2020; more than 50 billion IoT devices will be connected to the internet [6] (also shown in Fig. 2). This warning message also alarms about the emergence of new dangers, increased potential new misuses and cyber-crimes involving traditional criminal activities like fraud, theft, mischief and forgery. IoT development is based on evolving technologies like RFID, NFC, 3G & 4G communication technologies, wireless sensors and localization technologies due to which IoT is alleged of spying on people or misusing the personal collected information [7][8]. By leveraging these intelligent IoT technologies humans can be monitored, their security can be breached and their privacy can be infringed. With the adoption of these self-coordinating, and self-configuring smart IoT technologies, IoT users are facing ethical challenges. Thus, IoT has raised the opportunities for engaging in illegal or unethical behavior either consciously or unconsciously.

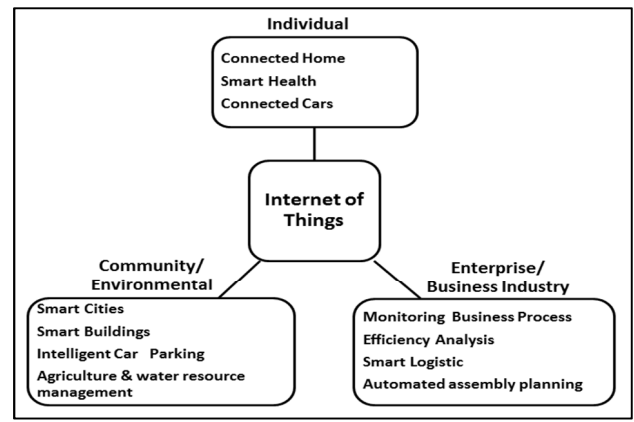

Fig. 1. Applications of Internet of Things (IoT)

Ethics is a branch of philosophy that defines the human conduct and behavior in the society. Ethics considers what is morally right or wrong, just or unjust, while rationally justifying our moral judgments. A number of research efforts have been done regarding the computer ethics [9][10], information era ethical issues [11][12] and IT related ethical issues [13][14]. The number of research efforts considering the relationship between computer or IT ethics and Islam is scant, receiving little or no attention [15-18]. A few researchers attempted to highlight some ethical issues specifically in IoTs [19]. According to PEW research center, Muslims were about 23.2% of the world population in 2010 and by 2050 this number is expected to increase up to 29.7% [20].

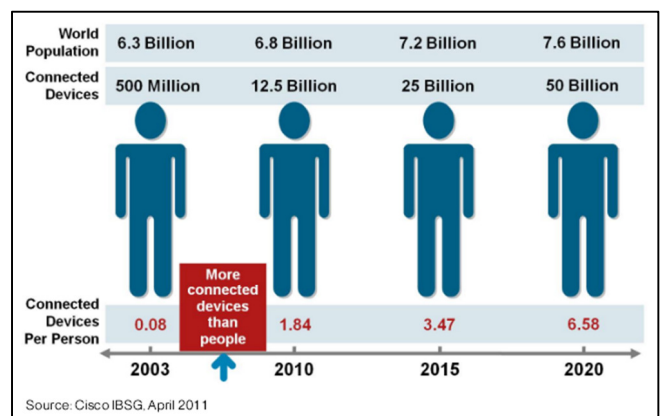

Fig. 2. Cisco IBSG estimates IoT [6].

Fig. 3, shows that Islam is the fastest growing religion and by 2050, Muslims are projected to increase by 73% with high fertility rates [20]. Since Muslims population in the world is getting higher, Muslims cannot remain isolated from the IoT hype. If the IoT vendors sand manufacturers consider the Islamic code of conduct for preserving their Muslim IoT user's security and privacy, the adoption of these IoT technologies among the Muslims will be significantly increased.

The researchers are continuously trying to propose technical solutions to handle ethical problems related to security and privacy concerns of the users which include data encryption, cyphering of radio communication or authentication techniques for communication devices, thus making it a bit difficult for unauthorized personnel to access important data or information. But still hackers and unauthorized persons use profiling and spying techniques to launch malicious attacks or misuse the information. Thus, these technical solutions still fail to solve some ethical problems in IoT.

There are some of the examples that clearly show that IoT applications have inflated serious ethical problems. For example in the Netherlands, smart electricity meters were planned to be introduced in order to optimize electrical power consumption and to reduce $CO_2$ emissions. But then Dutch parliament rejected the project because of the privacy problems created by such "spying devices" [21-23]. The technology of smart meters was already providing secure encryption of wireless data transmission but still the persons that can access the unencrypted data can use profiling to misuse the information. Smart baby monitors are normally used for monitoring the breathing, skin temperature, position and activities of infants. But some of these baby monitors are not safe from hackers and have security flaws and loopholes that can allow the hackers to remotely spy the babies and parents too [24].

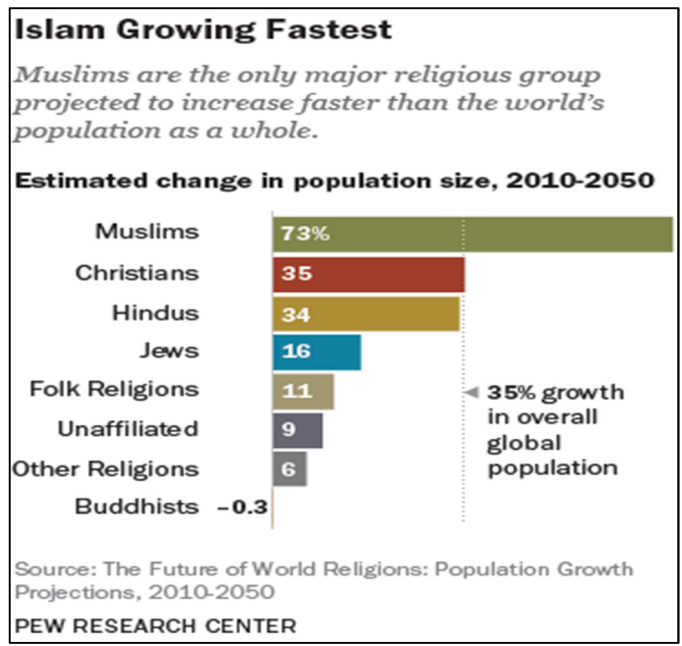

Fig. 3. The Future of World Religions [20]

Voice recognition capability of smart devices in IoT can be welcome addition but some Samsung Smart TV models can record personal conversations or sensitive information by capturing voice commands and associated text and transmit it to the third party. Although it is clarified by Samsung personnel that this feature is only used for improving the voice recognition, and these smart TVs are programmed to "wake up" when they detect a pre-programmed phrase such as "Hi TV," otherwise they remain in sleep mode, not storing or reporting anything. But, still there is chance that someone's personal sensitive information can be misused or private conversations can be shared [25], ultimately raising ethical issues. So whether it is smart thermostat, smart refrigerator or any other smart house hold device, there are many reports that these internet connected devices escalated unwanted invasions.

In-depth discussion regarding the ethical challenges raised by IoT is missing in the literature. Thus, the prime objective of this study is to highlight the ethical challenges that are upraised due to the adoption of these IoT technologies. We profoundly review the underlying Islamic concepts (Quranic verses) for using IoT technologies ethically.

The rest of the paper is organized as follows: Section 2 presents the ethical challenges of IoT. Section 3 shows how to deal with the ethical challenges of IoT using Islamic teachings. Section 4 finally concludes the paper.

## II. ETHICAL CHALLENGES OF IoTs

Digital convergence in IoT has raised complex ethical, legal and societal issues. Professional organizations like ACM, IEEE and ABET have established codes of ethics not only to overcome these issues and challenges but also to help IT professionals, specially end users in understanding and managing their ethical responsibilities. In this section we present some serious ethical challenges that are faced by the users of IoT technologies, including security and privacy the two major elements that lead to ethical challenges in IoT.

### A. Security

IoT smart devices (smart home appliances. connected cars, smart healthcare devices, smart factories and a lot others) are successful in providing the users ease of life and thus improving the quality of life but a large volume of sensitive information is transferred over IoT which pose great security risks. When the security in IoT is breached, the communication will be accessed, manipulated and fabricated by the unauthorized intruders who have no right to access that information. The smart IoT devices have limited CPU and processing power to incorporate existing encryption and authentication techniques and thus many such devices are at a sever security risk. For example, when a smart meter sends the information that power usage has dropped, this could indicate that a home is empty, making it an ideal target for a burglary or worse. We have also given other hacking examples in the introduction which makes it clear that IoT technology needs novel security solutions to handle such security breaches, since cybercriminals can potentially exploit this weakness of IoT for malicious purposes.

## B. Privacy

Privacy protection is the most essential element in IoT applications, since the hackers and unauthorized personnel can retrieve, explore, misuse, disclose or disseminate someone's personal information for bad intention that might be used to fake or fabricate the genuine information, or for embarrassment etc. Personal sensitive information captured from variety of IoT smart devices can also be used for committing identity theft, harassment or annoyance. Most IoT smart healthcare devices record their owner's personal information that is associated with ones' privacy (identity or location etc.) that can be abused by insurance companies or employers. The Health related data about the owners of smart IoT health care devices that one may not want to share or disclose can also be hacked from smart healthcare devices and can be misused which is stressful for the owners of these devices. Smart home appliances and others (e.g. connected car) which leverage RFID [26], GPS and NFC enabling technologies can be used to track the owner's location and movements form one place to another without the knowledge of the device owner and the location information can then be used for malicious purposes or might endanger one's life.

## III. QURANIC VERSES REGARDING ETHICAL ISSUES

All the major religious teachings including Islamic, Christianity and Judaism are not directly applicable on the issues related to security and privacy of today's technologies. But these religions teach their believers the guide lines to way of life. Particularly Islam highly emphasizes upon ethical values, involving every facet of human life. It is repeatedly stressed by many moral principles and codes of ethics throughout the holy Quran to observe ethical and moral code in human behavior. It is comprised of moral principles and direction to comprehend what is right and what is wrong. These Islamic teachings and principles are comprehensive and fair since it is historically proved that Islamic codes of conduct can successfully build ethically great societies and thus can be used to deal with new IoT technologies so as to apply them ethically correct. Muslims need to understand these Islamic principles and implement Islamic ethics in their usage of IoT technologies.

Almost all the aspects of life can be learned from Islamic principles that are beautifully explained in the Quran. Islamic principles teach us not to access the properties of others without their permission. Islamic teachings prohibit us from breaking any security system; it can be either some personal information or computer security system in IoT. Islam also places rewards for good deeds and good ethics and strict penalties and punishments for criminals and those who interrupt the security of others.

*"And those who harm believing men and believing women for [something] other than what they have earned have certainly born upon themselves a slander and manifest sin."* **Source: Quran (33:58).**

*"The Muslim is the one from whose tongue and hand the people are safe…"* **Source: Sunan al-Nasā'ī 4998.**

*"Whoever believes in Allah and the Last Day, let him not harm his neighbor."* **Source: Ṣaḥīḥ al-Bukhārī 6110.**

*"Whoever harms others, then Allah will harm him. Whoever is harsh with others, then Allah will be harsh with him."* **Source: Sunan al-Tirmidhī 1940.**

Islamic approach is that one cannot disclose the privacy of others. Islam teaches that the members of the society must respect others privacy.

*"O you who have believed, do not enter houses other than your own houses until you ascertain welcome and greet their inhabitants. That is best for you; perhaps you will be reminded."* **Source: Quran (24:27).**

*"O you, who have believed, avoid much [negative] assumption. Indeed, some assumption is sin. **And do not spy or backbite each other.** Would one of you like to eat the flesh of his brother when dead? You would detest it. And fear Allah; indeed, Allah is accepting of repentance and Merciful."* **Source: Quran (49:12).**

*"And it is not righteousness to enter houses from the back, but righteousness is [in] one who fears Allah. And enter houses from their doors. And fear Allah that you may succeed."* **Source: Quran (2:189).**

Form the above verses of Quran and hadith it is clear that ethical problems can be solved by following these Quranic verses and hadiths.

## IV. CONCLUSION

New developments and advancements in IoT may lead to most of the positive consequences like quality of human life is significantly improved. But on the dark side of IoT, it has given rise to some of the awful and unacceptable outcomes that has bad effects to the society, including illegal and immoral activities, social and cybercrimes, ultimately raising serious ethical issues. With the adoption of new IoT technologies some serious ethical issues will be raised. In this paper we have identifies these ethical issues and challenges and show how Islamic teachings can manage these challenges. In this concern, we have also quoted some of the most related verses of Quran and Hadith. By adopting and considering these significant verses, ethical problems can be minimized.